\documentstyle[prl,aps]{revtex}
\begin{document}
\wideabs{
\title{Spontaneous Parity Violation in QCD At Finite Temperature: On the Inapplicability of the Vafa-Witten Theorem}
\author{Thomas D.~Cohen}
\address{Department of Physics, University of Maryland, College Park, 
20742-4111\\
DOE/ER/40762-218~~~~UM PP \#01-033}
\date{\today}
\maketitle

\begin{abstract}
The generalization of the Vafa-Witten theorem ruling out parity violation to  QCD at finite temperature is considered.  It is shown that this generalization of the theorem rules out Lorentz-invariant parity violating operators from spontaneously acquiring vacuum expectation values.  However, it does not rule out Lorentz-noninvariant parity-violating operators from acquiring expectation values.  Other situations where the theorem is inapplicable are also discussed. \end{abstract}
\pacs{}}

More than 15 years ago Vafa and Witten constructed a very simple and elegant argument based on functional integrals that QCD cannot spontaneously break parity \cite{vw1}.  This so-called Vafa-Witten theorem has assumed increased importance over the years since the argument appears to be essentially unchanged when generalized to a finite box size in the temporal direction and hence immediately applies to QCD at finite temperature.  This is of potential significance in the physics of relativistic heavy ion physics since it implies that the high temperature phase of QCD cannot spontaneously break parity.  Thus suggestions that a high temperature parity violating phase might be created in heavy ion collisions\cite{pvp} are restricted to considerations of metastable phases or phases that occur at finite baryon density.  If there was no Vafa-Witten theorem, the possibility of a stable parity violating phase at high temperature would require serious consideration.  Proposed signatures in heavy ion collisions for a parity violating phase in QCD\cite{pvp} are applicable regardless of whether the phase is stable or metastable.

The purpose of the present letter is to clarify the status of the Vafa-Witten theorem for finite temperature.  It will be shown that the generalization of the Vafa-Witten theorem to finite temperature is nontrivial and that while the theorem forbids Lorentz invariant purely gluonic parity violating operators from acquiring expectation values, it does not imply that Lorentz-noninvariant parity-violating operators must have vanishing expectation values. Thus spontaneous parity violation at finite temperature has not been ruled out.
Additional situations where the theorem does not apply will also be discussed. In particular as noted in ref.~\cite{Aoki2}, the theorem does not generally apply to operators which contain quark bilinears.

Let us begin with a brief recapitulation of the Vafa-Witten argument. The problem is formulated in terms of a Euclidean space functional integral.  The expectation value of an operator, ${\cal O}$ is given by $\langle {\cal O} \rangle \, = \,  \, \frac{1}{V} \, \frac{\partial \log \left ( Z( \lambda ) \right )}{\partial \lambda}$ where $V$ is the volume of space-time; $Z(\lambda)$, the generating function, is given by 
\begin{equation}
Z(\lambda) \, = \, \int \, d[A] \,{\rm Det}[\not \! \!D - m] 
\, e^{- S_{\rm YM}} \, e^{- \int \, d^4 x   \lambda {\cal O}(x)}
\label{z} \end{equation} 
where $ {\cal S}_{\rm YM}$ is the action for the pure Yang-Mills theory, and the functional determinant is the result of integrating out the quarks.  

One key ingredient to the Vafa-Witten proof is that both the functional determinant and $e^{-S_{\rm YM}}$ are real and non-negative \cite{vw2}. The second key ingredient is that the only way to create a parity-violating Lorentz scalar operator out of gluons is by combining gluon fields and covariant derivatives with an odd number of the full anti-symmetric epsilon tensor.  An example of such an operator is $F_{\mu\nu} \tilde{F}^{\mu\nu} \, = \, 1/2 \epsilon^{\mu \nu \alpha \beta} F_{\mu \nu} F_{\alpha \beta}$.  Such a construction necessarily yields an operator that is purely imaginary when analytically continued from Minkowski space to Euclidean space.  Thus, the factor $ \, e^{ - \int \, d^4 x   \lambda {\cal O}(x)}$ must be replaced by
$ \, e^{ - i \int \, d^4 x   \lambda {\cal O}(x)}$  which is a pure phase.  Such a phase factor in an integration over a positive definite measure can only reduce the integral from what it would have been had the phase factor been unity; $Z(\lambda)$ is maximal for $\lambda = 0$. The generating function, $Z(\lambda)$, can be expressed as $Z(\lambda) = e^{- V {\cal E}(\lambda)}$ where ${\cal E}(\lambda)$ is the energy density for the system including the perturbation $-i \lambda {\cal O}$.  Thus ${\cal E}(\lambda)$ is a free energy and the fact that $Z(\lambda)$ is maximal at $\lambda=0$ implies that the free energy is a minimum. 

The free energy density near $\lambda=0$  is given by ${\cal E}(\lambda) = {\cal E}(0) + \lambda \langle {\cal O} \rangle_{\lambda=0}$.  If $\langle{\cal O} \rangle$ were to have a non-vanishing vacuum expectation value at $\lambda=0$, it can have either sign (since it is odd under parity which by hypothesis is broken) and thus regardless of the sign of $\lambda$ the theory can choose a vacuum state for which $\lambda \langle {\cal O} \rangle_{\lambda=0} <0$.  This implies that if $\langle {\cal O} \rangle_{\lambda=0} \ne 0$, then for small $\lambda$ there exists a state with lower free energy than for $\lambda=0$, in  contradiction to the general proof  that ${\cal E}(\lambda)$ is a global minimum at $\lambda=0$.  Thus $\langle{\cal O} \rangle=0$ for all purely bosonic parity violating operators.

Before proceeding, a few comments should be made.  The first is that the identification of ${\cal E}$ as a free energy is nontrivial.  The perturbation added to the system when including the $\lambda {\cal O}$ term is at the level of the Lagrangian and not the Hamiltonian.  Since the operator, ${\cal O}$ contains time derivatives the presence of the $\lambda {\cal O}$ term alters the definition of the canonical momenta in QCD.  For this reason ${\cal E}$ is not equal to $\langle {\cal H} + \lambda {\cal O} \rangle$ (where ${\cal H}$ is the canonical QCD Hamiltonian density(with $\lambda=0$).  Nevertheless ${\cal E}$ is clearly the appropriate free energy density in the sense that in Minkowski space $\langle {\cal O} \rangle \, = \, - \frac{\partial {\cal E}(\lambda)}{\partial \lambda}$. The fact that the free energy density is not $\langle {\cal H} + \lambda {\cal O} \rangle$ explains why  ${\cal E}(\lambda)$ is not a maximum at $\lambda=0$ in spite of the general result that a linear perturbation to the Hamiltonian for an operator with zero expectation value always lowers the energy.

Reference \cite{vw1} was a bit cavalier in its treatment of limits.   Symmetries cannot be spontaneously broken for finite systems.  Usually one considers a finite but spatially large system (of 3-volume $V_{\rm space}$) and then adds a symmetry-breaking perturbation to the system of strength $\lambda$.  The system spontaneously breaks provided that an intensive symmetry-breaking operator has a non-vanishing expectation value in the limit $\lambda \rightarrow 0$, $V_{\rm space} \rightarrow \infty$ with the infinite volume limit taken first.  Now the Vafa-Witten argument makes use of a small $\lambda$ limit {\it i.e.} a linear relation in the energy density and the source.  However, if the $\lambda \rightarrow 0$ and $V_{\rm space} \rightarrow \infty$ do not commute, it remains conceivable that the linear expression is not valid in the infinite volume limit.  For example, one might imagine that for small $\lambda$ and large $V_{\rm space}$ that the free energy density is of the following form:
\begin{equation}
{\cal E}(\lambda) = {\cal E}_0 + \sqrt{\alpha V_{\rm space}^{-\beta} + \gamma \lambda^2}
\label{straw}
\end{equation}
where $\alpha$, $\beta$ and $\gamma$ are positive constants.  Such a form has a minimum at $\lambda=0$ for any volume; in the infinite volume limit, however it develops a kink indicating spontaneous symmetry breaking.  The possibility of an energy density of the form of Eq.~(\ref{straw}) seems to indicate a loophole in the Vafa-Witten proof.  However, on physical grounds, one can rule this form out. In typical cases of spontaneous symmetry breaking the form analogous to Eq.~(\ref{straw}) (but with a minus sign) emerge as the lowest-lying state when a level crossing occurs at $\lambda=0$.  (For finite systems, the level crossing is avoided by a small interaction between the two vacua which give rise to the $\alpha V_{\rm space}^{-\beta}$ term in the square root.) While such a situation can easily arise at a relative maximum in ${\cal E}(\lambda)$ plainly it cannot at a global minimum since one of the levels has to pass under the other.  Thus, this apparent loophole in the Vafa-Witten argument is closed. 

Let us consider the generalization of this argument to finite temperature.  The functional determinant is simply replaced by a functional determinant over a finite spatial extent, $\tau$, which is the inverse temperature, and appropriate boundary conditions are imposed (periodic for gluons, anti-periodic for quarks).  The fermion determinant remains manifestly non-negative and hence the Vafa-Witten argument goes through unchanged: no Lorentz invariant parity violating operator can acquire a vacuum expectation value.  However, this does not rule at parity violation at finite temperature even for pure gauge theory. 

The heat bath has a rest frame and thereby violates Lorentz symmetry. This introduces a 4-vector, the 4-velocity $u_{\mu}$ into the problem. Thus, at finite temperatures, it is possible for Lorentz non-invariant operators to have non-vanishing expectation values.  Moreover, unlike the Lorentz-invariant parity-violating operators considered by Vafa and Witten one can easily find Lorentz-noninvariant parity-violating scalar operators which do not pick up a factor of $i$ when analytically continuing from Minkowski to Euclidean space. Eamples of such operators include ${\rm tr}[(\vec{D} \times \vec{B})  \cdot \vec{B}]$ and ${\rm tr}[(\vec{D} \times \vec{E})  \cdot \vec{E}]$ where the trace is over color.  The forms for these operators clearly are not manifestly covariant.  However, such operators can be expressed covariantly in terms of $F_{\mu \nu}$ and the four-velocity of the heat bath. For example in the rest frame of the heat bath, ${\rm tr}[(\vec{D} \times \vec{E})  \cdot \vec{E}] =
\epsilon^{\alpha \beta \gamma \delta} {\rm tr}[ D_{\alpha} F_{\gamma \epsilon} F_{\delta \phi} ] u_\beta u^\epsilon u^\phi$. Since these operators are purely real in Euclidean space one cannot show that their inclusion as a source must decrease the generating function.  Thus the Vafa-Witten theorem does not apply for such operators and one cannot rule out the possibility that these operators spontaneously acquire expectation values at finite temperature.

Apart from the generalization to finite temperature, the validity of the Vafa-Witten theorem has recently been called into doubt.  The basic argument of Vafa and Witten described above is cast in terms of expectation values of purely bosonic parity-violating Lorentz scalar operators.  In a footnote, they argue that the analysis generalizes to operators containing quark bilinears. This generalization has been questioned in connection with the so-called Aoki phase of lattice regularized QCD with Wilson fermions at non-zero lattice spacing\cite{Aoki} which breaks both parity and flavor.  While the Aoki phase is clearly a lattice artifact, the assumptions used to derive the Vafa-Witten theorem appear to be valid in lattice QCD with Wilson fermions.   It is clear that the loophole allowing for the Aoki phase is fairly general.  (As will be discussed below, examples can be found where continuum QCD exhibits parity violation.)  As discussed in Ref. \cite{Aoki2} this loophole can be traced to the generalization of the theorem to operators including quarks.

Consider the case of parity-violating scalar operators that include fermion bilinears.  In the original Vafa-Witten paper, it is argued that when one integrates out the fermions the operator becomes a purely bosonic operator and the previous argument applies\cite{vw1}.  Thus, for example, if ${\cal O}=\overline{q} i \gamma_5 q$, then  upon integrating out the quarks ${\cal O}$ becomes ${\rm tr}[i \gamma_5 S_A(x,x) ]$ where $S_A$ is a quark propagator in the presence of the gluon field configuration. This operator is a functional of $A$ only.  

However, there is clearly something wrong with this argument. There are several counterexamples for which the functional integral measure is strictly nonnegative but parity-violating operators constructed from fermion bilinears have non-vanishing vacuum expectation values.  One case is QCD with strictly zero quark masses for two flavors.  In this case, the theory has no explicit chiral symmetry breaking.  Spontaneous chiral symmetry breaking means, however, that a source of the form $\lambda \overline{q} i \gamma_5 \tau_a q$ will cause a vacuum expectation value for $ \overline{q} i \gamma_5 \tau_a q$, even in the limit $\lambda \rightarrow 0$.  A second counterexample is QCD at finite  isospin density\cite{iso}.  If an isospin violating chemical potential term, $\mu \overline{q} \gamma_0 \gamma_5 \tau_3 q$ is added to the QCD Lagrangian, then for $|\mu | > m_{\pi}$, the system will acquire an s-wave pion condensate so that $\langle \overline{q} i \gamma_5 \tau_{1,2} q \rangle$ is nonzero.  This occurs despite the fact that with a finite chemical potential, the fermion determinant is real and nonnegative \cite{iso2}.  A third example is the Aoki phase mentioned above.  

The problem with the extension of the Vafa-Witten theorem to operators containing fermion bilinears has been identified in connection with the Aoki phase\cite{Aoki2}.  The bosonic operator obtained after integrating out the fermions is either not purely imaginary or is ill defined\cite{Aoki2}.  More generally, it is easy to see that the addition of sources containing quark bilinears and of strength $\lambda$ do not necessarily imply that $Z(\lambda)$ is a maximum at $\lambda=0$ but is often a minimum. The same considerations invalidate the theorem for the other counterexamples.   Consider, for example two-flavor QCD at zero quark mass with a perturbation of the form $\lambda \overline{q} i \gamma_5 \tau_a q$.  Upon integrating out the quarks, the effect on the functional integral of the inclusion such a perturbation is completely contained in the functional determinant. It becomes ${\rm Det}[\not \! \! \! D - i \lambda \gamma_5 \tau_a]$, which may be expressed naturally in terms the eigenvalues of the $\, \not \! \! \!D$ operator.  It is simple to see that if $\psi_j$ is an eigenstate of $ \, \not \! \! \!D $ with eigenvalue $i \epsilon_j$ then $\tau_a \gamma_5 \psi_j$ is also an eigenstate but with eigenvalue of $- \epsilon_j$.  From this, it is straightforward to write the functional determinant as
\begin{equation}  
 {\rm Det}[\not \! \!D - i \lambda \gamma_5 \tau_a] \, = \, 
\prod_{j} (\epsilon_j^2 + \lambda^2)
\label{det}
\end{equation}
where the product is evaluated over a single flavor and the effect of two flavors is included in the square.  From the form of Eq.~(\ref{det}) it is immediately
apparent that  ${\rm Det}[\not \! \!D - i \lambda \gamma_5 \tau_a]$ has a minimum at $\lambda=0$ which in turn implies that $Z(\lambda)$ has a minimum at $\lambda=0$.  The situation for QCD at finite isospin density is analogous.  It is not surprising that the inclusion of fermion bilinears gives a $Z(\lambda$) with a maximum at $\lambda=0$ rather than a minimum. Unlike in the  case of purely bosonic operators, perturbations proportional to the parity violating fermion bilinears considered here do not alter the definition of the conjugate momenta.  Thus, they act as linear perturbations on the Hamiltonian as well as on the Lagrangian and, hence, by very generally variational arguments always lower the free energy density.

The Vafa-Witten theorem has also been questioned at a more fundamental level.   
Azcoiti and Galante recently argued that the Vafa-Witten proof implicitly depends on the existence of a well-defined free-energy density for small external parity violating perturbations, and that this implicit assumption is, in fact, not true in the presence of spontaneous parity violation \cite{AG}. 
As the thermodynamic limit plays a central role in the analysis, the four-volume $V$ will be explicitly denoted throughout.  The key theoretical tool used in their analysis is the probability distribution function for the operator ${\cal O}$ defined as
\begin{eqnarray}
& & P(\tilde{\cal O},V) = \nonumber \\
 & & Z^{-1}(0, V) \, \int \, d[A] \,{\rm Det}[\not \! \!D - m] \, e^{- S_{\rm YM}} \, \delta(\overline{\cal O}(A) - \tilde{\cal O})
\label{pdf} 
\end{eqnarray} 
where $\overline{\cal O}(A) \equiv V^{-1} \int \, d^4 x \, {\cal O}$ and $Z(\lambda, V)$ is the generating function.  By construction $P(\tilde{\cal O},V)$ is the probability that a configuration removed at random from the functional integral will have $\overline{\cal O}(A) = \tilde{\cal O}$.   The generating functional $Z(\lambda, V)$ may be expressed as an
integral over $P(\tilde{\cal O},V)$:
\begin{equation}
Z(\lambda, V) \, = \,Z(0,V) \,  \int \, d \, \tilde{\cal O} P(\tilde{\cal O},V) e^{- i\lambda V \tilde{\cal O}}  
\label{form}\end{equation}

 ${\cal O}$ is an intensive operator; in the thermodynamic limit, the probability distribution function will develop a delta function.  If the symmetry is broken, two delta functions will develop; one at positive value and one at negative value.  Denoting the expectation value of the operator in the symmetry broken phase as $a$, one has
\begin{equation}
\lim_{V \rightarrow \infty} P(\tilde{\cal O},V)  \, = \, \frac{1}{2} \delta(\tilde{\cal O} \, - \, a) \, + \, \frac{1}{2} \delta(\tilde{\cal O} \, + \, a) 
\label{delform}
\end{equation}
Azcoiti and Galante show that whenever the symmetry is spontaneously broken in the sense of  Eq.~(\ref{delform}), $Z(\lambda,V )$ passes from positive to negative an infinite number of times as $\lambda$ is varied at fixed $V$. This in turn implies that ${\cal E}= -\frac{1}{V} \log \left ( Z(\lambda, V) \right )$ is not well defined since the log has multiple branches.

In Ref.~\cite{AG} it was argued that the nonexistence of a well-defined free energy density for the case of spontaneous symmetry breaking invalidates the proof of the Vafa-Witten theorem.   The proof implicitly makes use of the existence of a well-defined free energy (for example when writing ${\cal E}(\lambda) = {\cal E}(0) + \lambda \langle {\cal O} \rangle_{\lambda=0}$); the nonexistence of well-defined free energy apparently invalidates the proof of the Vafa-Witten theorem. However, the argument of Ref.~\cite{AG} can be turned on its head: since Azcoiti and Galante have shown that spontaneous symmetry breaking in the sense of Eq.~(\ref{delform}) implies an ill-defined free energy,  the existence of a well-defined free energy is equivalent to proving that there is no symmetry breaking.  It may be argued that the assumption of a well-defined free energy is quite innocuous, in which case it trivially follows that such operators cannot spontaneously break parity.  Whether this assumption is generically true is a delicate technical question that will not be addressed here.  It should be noted, however, that the inapplicability of the Vafa-Witten theorem to Lorentz non-invariant operators at finite temperature, and to operators constructed from quark bilinears discussed above, is independent of the validity of the assumption of a well-defined free energy. 

In summary, it has been shown that the Vafa-Witten theorem does not generally rule out parity breaking at finite temperature.  Parity violating but Lorentz non-invariant operators are not precluded from spontaneously acquiring expectation values at finite temperatures.  Of course, the fact the Vafa-Witten theorem does not forbid parity breaking at finite temperatures does not mean that it is likely to occur.  On the whole, one expects that parity breaking at finite temperature to be rather unlikely since generically one expects a high temperature phase to be less ordered than a low temperature phase.  However, QCD is in many respects a quite unusual theory and the possibility of parity breaking at finite temperature should be seriously  considered.

Discussion with S. Nussinov, M. Stephanov and V. Azcoiti are gratefully acknowledged.  This work was supported by the U.S.~Department of Energy via grant DE-FG02-93ER-40762.

\end{document}